\documentclass{aa520}
\usepackage{graphicx}
\usepackage{txfonts}
\newcommand{\epsindi}{$\varepsilon$\,Indi}
\newcommand{\epsindia}{$\varepsilon$\,Indi\,A}
\newcommand{\epsindib}{$\varepsilon$\,Indi\,B}
\newcommand{\epsindiba}{$\varepsilon$\,Indi\,Ba}
\newcommand{\epsindibb}{$\varepsilon$\,Indi\,Bb}
\newcommand{\epsindibab}{$\varepsilon$\,Indi\,Ba,Bb}
\newcommand{\degree}{\mbox{$^{\circ}$}}               
\newcommand{\micron}{\mbox{\,${\mu}$m}}               
\newcommand{\Lsolar}{\mbox{\,$L_{\odot}$\/}}          
\newcommand{\Rsolar}{\mbox{\,$R_{\odot}$\/}}          
\newcommand{\Mjup}{\mbox{\,$M_{\rm Jup}$\/}}          
\newcommand{\magnit}[2]{\mbox{$\mbox{\rm #1}^{\mbox{\rm\tiny m}}%
     \!\!\!.\!\,\, \mbox{\rm #2}$}}                   
\newcommand{\magap}[1]{\mbox{$\mbox{\rm #1}^{\mbox{\rm\tiny m}}$}} 
\newcommand{\oversim}[2]{\lower0.5ex\vbox{\baselineskip=0pt\lineskip=0.2ex
     \ialign{$\mathsurround=0pt #1\hfil##\hfil$\crcr#2\crcr\sim\crcr}}}
\newcommand{\etal}{\mbox{et al.}}         
\newcommand{\eg}{\mbox{\hbox{e.g.,}}}             
\newcommand{\idest}{\mbox{\hbox{\it i.e.,}}}          
\newcommand{\cf}{\mbox{\hbox{\it cf.}}}               
\newcommand{\kmpers}{\mbox{\,km\,s$^{-1}$}}           
\begin{document}
\title{\epsindibab: the nearest binary brown dwarf%
\thanks{Based on observations collected with the ESO VLT, Paranal, Chile.} 
}

\author{M.~J.~McCaughrean\inst{1} \and
        L.~M.~Close\inst{2} \and
        R.-D.~Scholz\inst{1} \and
        R.~Lenzen\inst{3} \and
        B.~Biller\inst{2} \and
        W.~Brandner\inst{3} \and
        M.~Hartung\inst{4} \and
        N.~Lodieu\inst{1}
}

\offprints{M. J. McCaughrean; {\tt mjm at aip dot de}}

\institute{$^1$Astrophysikalisches Institut Potsdam, 
           An der Sternwarte 16, 14482 Potsdam, Germany \\
           $^2$Steward Observatory, University of Arizona,
           933 N. Cherry Ave., Tucson, AZ 85721--0065, USA \\
           $^3$Max-Planck-Institut f\"ur Astronomie, K\"onigstuhl 17,
           69117 Heidelberg, Germany \\
           $^4$European Southern Observatory, Alonso de Cordova 3107,
           Vitacura, Santiago, Chile \\
}

\date{Received ...; accepted ...}

\titlerunning{\epsindibab: the nearest binary brown dwarf}
\authorrunning{McCaughrean \etal{}}

\abstract{
We have carried out high angular resolution near-infrared imaging and 
low-resolution (R$\sim$1000) spectroscopy of the nearest known brown dwarf, 
\epsindib, using the ESO VLT NAOS/CONICA adaptive optics system. We find it 
to be a close binary (as also noted by Volk \etal{} 2003), with an angular 
separation of 0.732 arcsec, corresponding to 2.65\,AU at the 3.626\,pc 
distance of the \epsindi{} system.
In our discovery paper (Scholz \etal{} 2003), we concluded that \epsindib{} 
was a $\sim$\,50\Mjup{} T2.5 dwarf: our revised finding is that the two 
system components (\epsindiba{} and \epsindibb) have spectral types of T1 
and T6, respectively, and estimated masses of 47 and 28\Mjup, respectively, 
assuming an age of 1.3\,Gyr. Errors in the masses are $\pm$10 and $\pm$7\Mjup,
respectively, dominated by the uncertainty in the age determination 
(0.8--2\,Gyr range). This uniquely well-characterised T dwarf binary system 
should prove important in the study of low-mass, cool brown dwarfs. The two 
components are bright and relatively well-resolved: \epsindib{} is the only 
T dwarf binary in which spectra have been obtained for both components. The
system has a well-established distance and age. Finally, their orbital motion 
can be measured on a fairly short timescale (nominal orbital period 
$\sim$\,15\,yrs), permitting an accurate determination of the true total 
system mass, helping to calibrate brown dwarf evolutionary models.
}
\maketitle

\section{Introduction}
Binary systems are a common product of the star formation process and there
is no reason to suspect that the same would not hold below the hydrogen-burning 
limit, in the domain of brown dwarfs. Binary systems offer important 
advantages in studies of the characteristics of brown dwarfs, in part because 
the two components are expected to be coeval and have the same chemical 
composition. The differential measurement of physical parameters such as
luminosity, effective temperature, and surface gravity would then provide
crucial constraints on evolutionary models. Furthermore, sufficiently tight
binary systems would allow the direct measurement of the system mass via
the monitoring of orbital motion.

After a number of unsuccessful searches, the first spatially-resolved brown
dwarf binary was found in the solar neighbourhood by Mart\'{\i}n,
Brandner, \& Basri (1999) and, subsequently, high spatial resolution imaging 
has identified
a significant number of such systems (\eg{} Close \etal{} 2002a; Goto
\etal{} 2002; Potter \etal{} 2002; Gizis \etal{} 2003; Close \etal{} 2003).
Indeed, roughly 20\% of a magnitude-limited sample of $\sim$135 L dwarfs and 
10 T dwarfs imaged with the HST have been shown to have candidate companions 
at projected separations of 1--10\,AU (Reid \etal{} 2001; Bouy \etal{} 2003; 
Burgasser \etal{} 2003). Many of these sources have since been confirmed as 
physical pairs with second epoch data.

Particularly important among brown dwarfs are those with well-established 
distances and ages, as their physical parameters can be accurately determined 
and they can be used as key templates in the understanding of the physical 
evolution of these substellar sources (\eg{} Gl\,229\,B: Nakajima \etal{}
1995; Gl\,570\,D: Burgasser \etal{} 2000).  An especially rewarding
discovery would be a binary brown dwarf system with a well-established 
distance and age, a small separation such that its orbit could be measured 
on a reasonable timescale, and yet nearby enough that its components would
be bright, well-resolved, and thus readily amenable to observations.

Scholz \etal{} (2003; hereafter SMLK03) recently reported the discovery of a 
new benchmark brown dwarf, \epsindib, as a very wide ($\sim$1500\,AU) 
companion to the nearby, very high proper-motion ($\sim$4.7\,arcsec/yr) 
southern star, \epsindia. With an accurate Hipparcos distance to the system 
of 3.626\,pc (ESA 1997), \epsindib{} was the nearest known brown dwarf to 
the Sun and the brightest member of the T dwarf class by roughly 2 magnitudes 
in the near-IR\@. In addition, through its association with \epsindia, it had 
a reasonably well-determined age of $\sim$1.3\,Gyr (likely range 0.8--2\,Gyr; 
Lachaume \etal{} 1999). This fortuitous combination of parameters made it 
a thus far unique object for detailed, high precision studies; in particular, 
high resolution spectroscopy of its atmosphere could be important as, with 
a spectral type of T2.5, it was one of the few objects in the transition 
zone between L and T dwarfs where dramatic changes in the atmospheric
properties are known to occur.

Another exciting prospect raised by the discovery of \epsindib{} was that
deep, high angular resolution imaging might reveal lower-mass companions, 
potentially even into the planetary domain, at separations small enough 
($\sim$\,1\,AU) that the orbit could be traced out in only a few years,
leading to an accurate, model-independent determination of the total system 
mass. 


We observed \epsindib{} with the NAOS/CONICA (henceforth NACO) near-IR adaptive 
optics system on UT4 (Yepun) of the ESO VLT, Paranal, Chile, on August 13 2003 
(UT). It was readily resolved into two components (henceforth \epsindiba{} and 
\epsindibb{} following the IAU-approved Washington Multiplicity Catalog 
nomenclature of Hartkopf \& Mason 2003), as also noted five days later by 
observers at the Gemini-South telescope (Volk \etal{} 2003). Here we present 
the first $\sim$0.1 arcsec resolution near-IR imaging and spectroscopy of the 
\epsindibab{} system, from which we determine accurate positions and spectral 
types for the two components. We then derive effective temperatures and 
luminosities, and make estimates of the masses based on evolutionary models.

\section{Imaging observations}
Adaptive optics imaging observations of low-mass stars and brown dwarfs
generally use the source itself for self-guiding and correction (\cf{} 
observations of L dwarfs by Close \etal{} 2003). However, as a T dwarf,
\epsindib{} presents a real challenge to adaptive optics. Despite its 
proximity, it is a very cool, low-luminosity object and too faint in the 
optical ($I$$\sim$\magap{17}) for accurate wavefront sensing. Fortunately 
though, it is significantly brighter in the near-IR ($K$$\sim$\magap{11}) and 
thus perfectly suited to the unique infrared wavefront sensing capability 
(IR WFS) of the NACO system (Lenzen \etal{} 2003; Lagrange \etal{} 2003).

\begin{table}
\caption[]{Relative astrometry and photometry for \epsindiba{} and \epsindibb.
The image scale in the NACO S27 camera was measured as $27.07\pm 0.05$
mas/pixel during NACO commissioning using astrometric binaries and Galactic
Centre imaging. The system position angle offset was measured using images
taken of the astrometric binary WDS\,19043$-$2132 on August 13 in the same
camera configuration: the error in the determination was $\pm$0.143\degree.
The separation and position angle were confirmed in independent data taken
in another NACO mode, for which the image scale was measured using both
WDS\,19043$-$2132 and a non-astronomical opto-mechanical setup. The errors 
in the mean separation and position angle include the NACO data measurement 
errors and the errors in the system parameters. Differential magnitudes 
measured by Volk \etal{} (2003) and Smith \etal{} (2003) at other wavelengths 
are also given for convenience.
}
\label{tab:nacoastphot}
\begin{center}
\begin{tabular}{ll}
\hline
\multicolumn{2}{c}{Separation (arcsec)} \\ \hline
$J$   & $0.733$ \\
$H$   & $0.732$ \\
$K_s$ & $0.731$ \\
mean  & $0.732\pm 0.002$ \\ \hline
\multicolumn{2}{c}{Position angle (\degree{} E of N)} \\ \hline
$J$   & 136.81 \\
$H$   & 136.78 \\
$K_s$ & 136.83 \\
mean  & $136.81\pm 0.14$ \\ \hline
\multicolumn{2}{c}{$\Delta$\,mag (Bb $-$ Ba) (NACO)} \\ \hline
$J$   & \magnit{0}{94}$\pm$\magnit{0}{02} \\
$H$   & \magnit{1}{76}$\pm$\magnit{0}{02} \\
$K_s$ & \magnit{2}{18}$\pm$\magnit{0}{03} \\ \hline
\multicolumn{2}{c}{$\Delta$\,mag (Volk \etal{} 2003)} \\ \hline
$I$            & \magnit{1}{5} \\
1.083\micron{} & \magnit{0}{69} \\
1.282\micron{} & \magnit{0}{47} \\
1.556\micron{} & \magnit{1}{34} \\
2.106\micron{} & \magnit{1}{92} \\
2.321\micron{} & $>$\magnit{3}{8} \\ \hline
\multicolumn{2}{c}{$\Delta$\,mag (Smith \etal{} 2003)} \\ \hline
1.647\micron{} & \magnit{1}{9} \\
2.321\micron{} & $>$\magap{3} \\
\hline
\end{tabular}
\end{center}
\end{table} 

At the time of our observations, the natural seeing was a very good $\sim$0.5
arcsec FWHM\@. To further ensure the best possible image correction, we 
used the N90C10 dichroic in NACO, which sends just 10\% of the source flux 
to the science camera, while diverting the remaining 90\% to the IR WFS,
which was run in 49 subaperture mode. Combined, these factors enabled us 
to obtain the sharpest ever (0.08 arcsec FWHM at 2\micron) infrared images
of a binary T dwarf.

Standard near-IR AO observing procedures were followed. In each of the $J$, 
$H$, and $K_s$ broad-band images, a total of 18 spatially dithered ($\sim$3 
arcsec) images were obtained, each with a 5 second integration time. The S27 
camera was used with an image scale of 27.07$\pm$0.05 milliarcsec/pixel and a 
total field-of-view of 27.7$\times$27.7 arcsec. The VLT derotator maintained 
north along the Y-axis of the science detector to within 
$0.06\pm 0.143$\degree{} throughout. Sky subtraction, flat-fielding, 
cross-correlation, and image alignment to within 0.01 pixels were carried out 
using custom scripts (Close \etal{} 2002a,b) and standard IRAF programs.
The total integration time in each filter was 90 seconds and the final 
FWHM was 0.116, 0.100, and 0.084 arcsec at $J$, $H$, and $K_s$, respectively.

\begin{figure*}
\centering
\includegraphics[width=5.9cm]{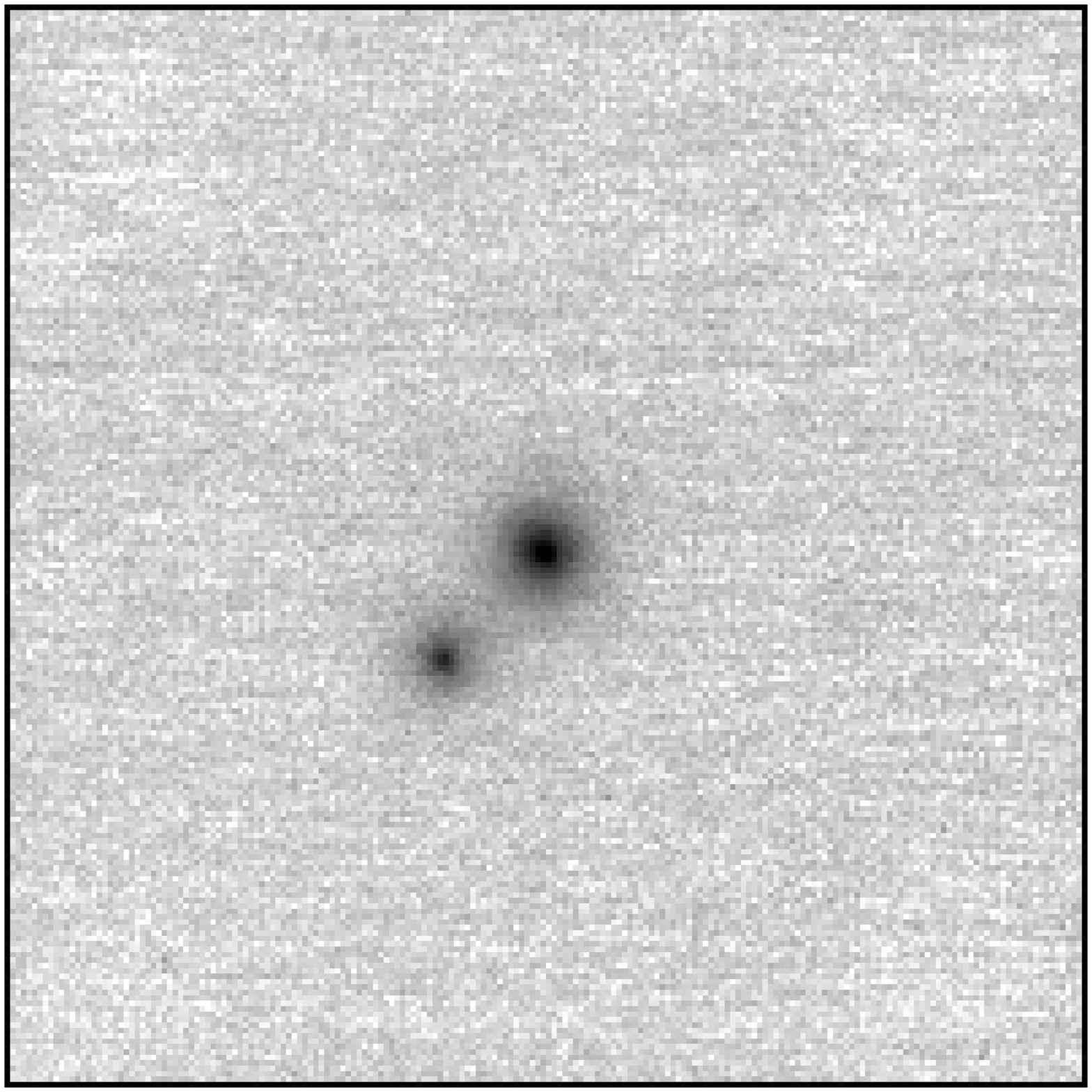} 
\includegraphics[width=5.9cm]{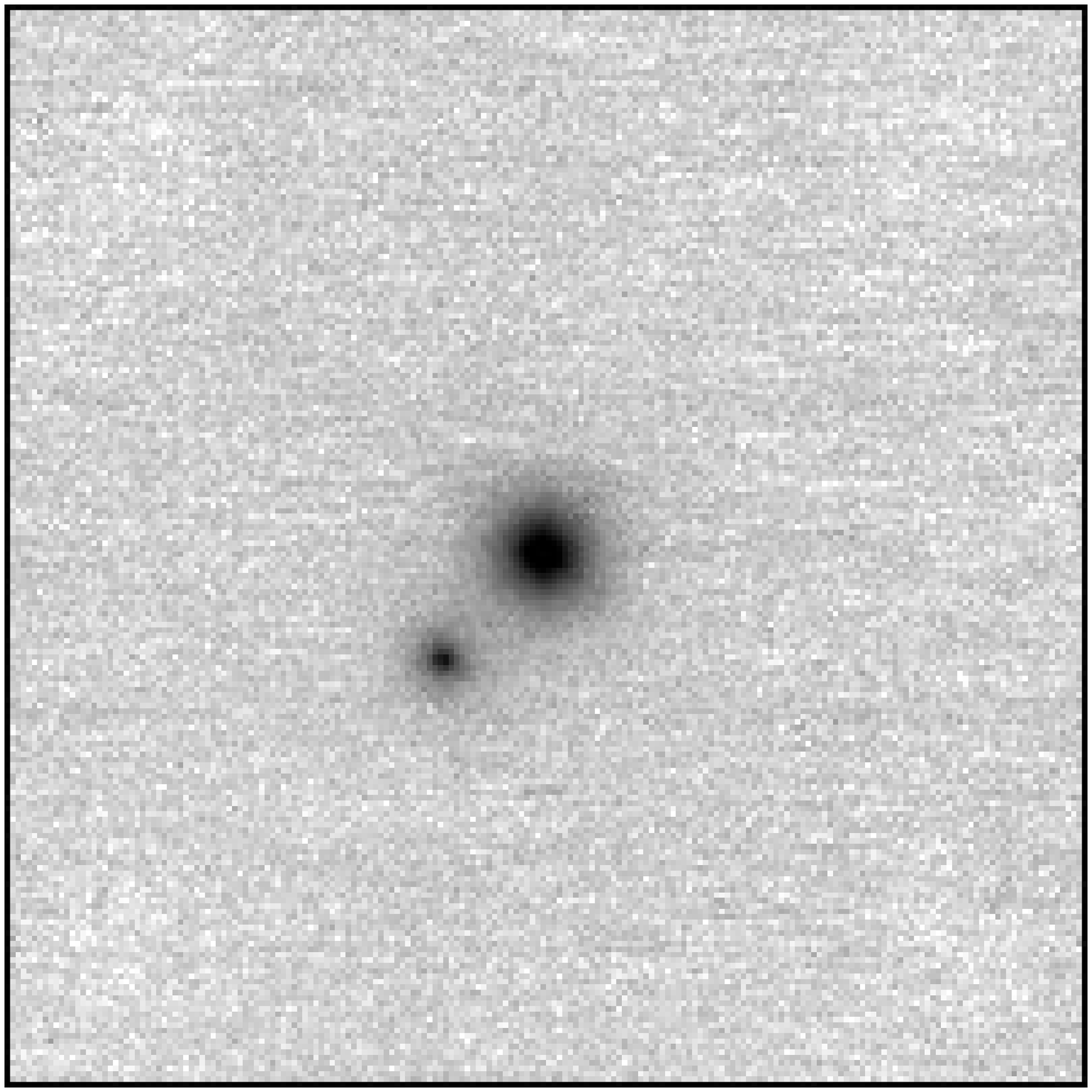} 
\includegraphics[width=5.9cm]{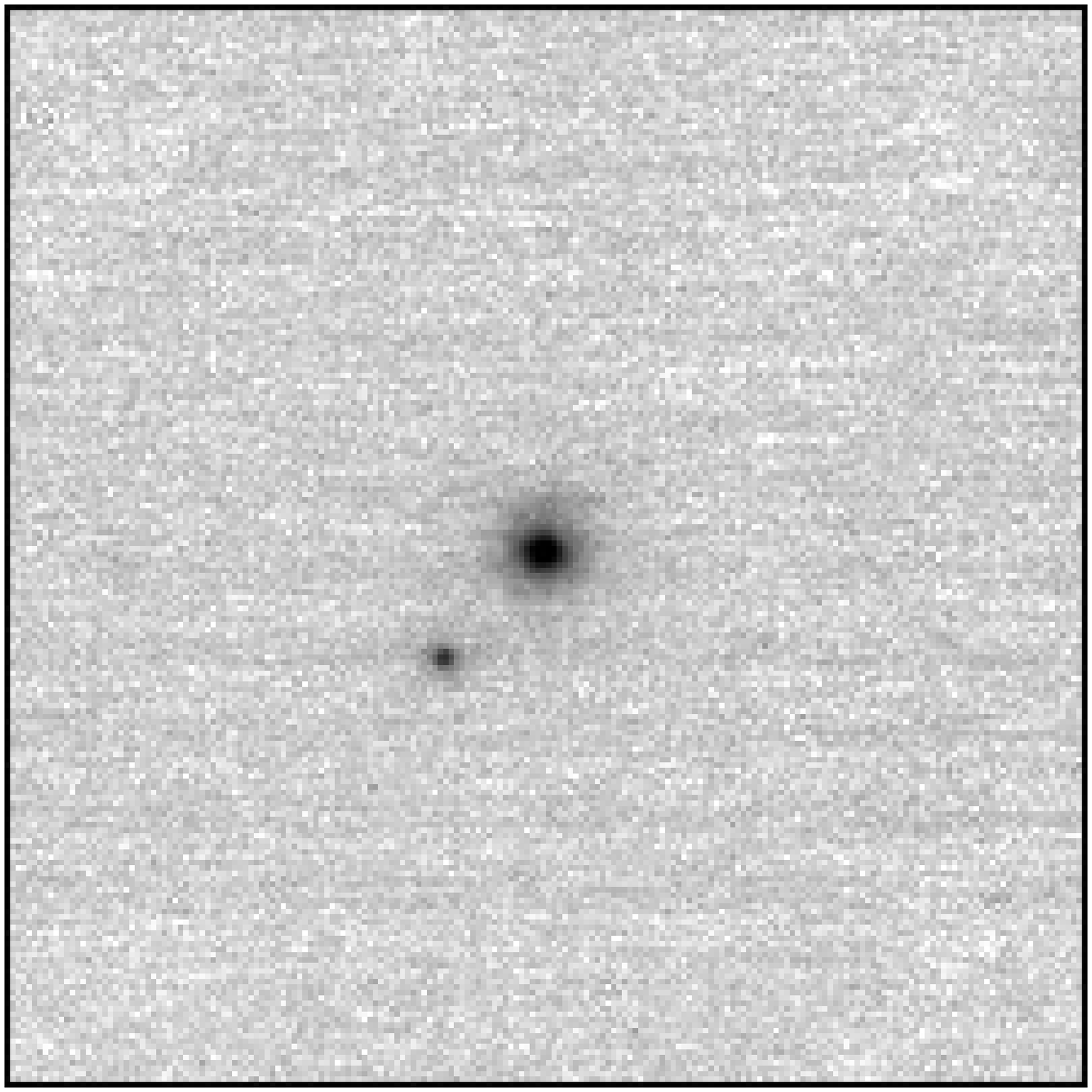}
\caption{NACO broad-band near-IR adaptive optics images of the 
\epsindibab{} system, with $J$, $H$, and $K_s$ from left to right. Each image 
is a 5.4$\times$5.4 arcsec (19.6$\times$19.6\,AU at 3.626\,pc) subsection of 
the full 27.7$\times$27.7 arcsec NACO S27 camera field-of-view. The angular
resolutions are $\sim$\,116, 100, and 84 mas FWHM at $J$, $H$, and $K_s$,
respectively. North is up, East left: \epsindibb{} is the fainter source to 
the south-east. The intensities are displayed logarithmically. No other 
sources are detected in any filter across the whole NACO FOV\@. 
}
\label{fig:nacoimages}
\end{figure*}

The resulting images are shown in Figure~\ref{fig:nacoimages}: the new, fainter
source, \epsindibb, is well resolved for the first time from \epsindiba, and 
is relatively bluer. The rest of the NACO field is empty to a limiting 
magnitude $\sim$\magap{3} fainter than \epsindibb, immediately suggesting that 
they constitute a physical pair. Volk \etal{} (2003) checked the 1999.9 epoch 
2MASS survey images at the coordinate where the fast-moving \epsindiba{} was 
located at the 2003.6 epoch of their Gemini-South observations. They found 
no source with the characteristics of \epsindibb, and thus concluded that the 
two sources must be comoving. Finally, we have checked some short, 0.6 arcsec
FWHM seeing VLT FORS1 optical acquisition images taken of \epsindib{} on June 
2003 which also in retrospect show the source to be binary. The separation 
and position angle of the system were measured to be 0.57$\pm$0.08 arcsec and 
139.5$\pm$2.6\degree, respectively, \idest{} close to the parameters measured
from the NACO images (Table~\ref{tab:nacoastphot}). Due to its very high
proper motion, \epsindib{} moved 0.8 arcsec in those two months: the 
separation and PA could not have remained constant unless \epsindibb{} 
is a comoving, physical companion to \epsindiba.

PSF-fitting photometry techniques (IRAF DAOPHOT) were used to measure the 
relative positions, position angle, and fluxes in all three filters
(Table~\ref{tab:nacoastphot}). The mean separation of $0.732\pm 0.002$ 
arcsec at the distance of $3.626\pm 0.01$\,pc distance of \epsindibab{} 
(SMLK03) corresponds to a projected spatial separation of 
$2.65\pm 0.01$\,AU\@. 

Table~\ref{tab:posphot} lists optical/near-IR photometry and positional data 
for the system taken from archival photographic plates, the 2MASS and DENIS 
near-IR sky surveys. There is good agreement between the three $K_s$/$K$ band 
magnitudes, but there is a relatively large difference ($\sim$\magnit{0}{2}) 
between the 2MASS Point Source Catalog (2MPSC) and DENIS $J$ magnitudes, and 
between the 2MPSC and 2MASS Quick Look Atlas (2MQLA) magnitudes derived 
independently by SMLK03 using M dwarfs for calibration. Also, the 2MPSC $H$ 
magnitude $\sim$\magnit{0}{3} is brighter than that derived from the 2MQLA by 
SMLK03. 

\begin{table}
\caption[]{Astrometry and photometry for the combined \epsindibab{} system
from the SuperCOSMOS Sky Surveys based on ESO Schmidt (SSS-ESO) and UK
Schmidt (SSS-UK) photographic plates (Hambly \etal{} 2001a,b), the 2MASS 
Quick Look Atlas (2MQLA; SMLK03), the 2MASS Point Source Catalog (2MPSC; 
Cutri \etal{} 2003), the DENIS second data release 
({\tt http://vizier.u-strasbg.fr/viz-bin/Cat?B/denis}), and a public
VLT FORS1 $I$ band acquisition image from another group (M\'enard, Delfosse,
\& Monin; ESO program 72.C-0575(A)).
}
\label{tab:posphot}
\begin{center}
\begin{tabular}{@{}ccrl@{}}
\hline
$\alpha, \delta$ (J2000.0)       & Epoch    & Magnitude   & Data    \\ 
\hline
$22~04~03.113$~~$-56~46~19.46$   & 1984.555 & $R$=20.75   & SSS-ESO \\
$22~04~09.465$~~$-56~46~52.58$   & 1997.771 & $I$=16.59   & SSS-UK  \\
$22~04~10.392$~~$-56~46~57.29$   & 1999.666 & $I$=16.77   & SSS-UK  \\ 
$22~04~10.52$~~~~$-56~46~57.8~~$ & 1999.855 & $J$=12.11   & 2MQLA   \\
                                 &          & $H$=11.59   & 2MQLA   \\
                                 &          & $K_s$=11.17 & 2MQLA   \\
$22~04~10.52$~~~~$-56~46~57.7~~$ & 1999.855 & $J$=11.91   & 2MPSC   \\
                                 &          & $H$=11.31   & 2MPSC   \\
                                 &          & $K_s$=11.21 & 2MPSC   \\
$22~04~10.97$~~~~$-56~47~00.4~~$ & 2000.781 & $I$=16.90   & DENIS   \\
                                 &          & $J$=12.18   & DENIS   \\
                                 &          & $K$=11.16   & DENIS   \\
$22~04~12.278$~~$-56~47~06.58$   & 2003.488 &             & FORS1   \\
\hline
\end{tabular}
\end{center}
\end{table} 

Although the 2MPSC and 2MQLA magnitudes are, in principle, derived from the
same source data, there are calibration issues with the Quick Look Atlas data
that may be responsible for the differences. Equally, it is well known that
there are often significant differences in the photometry of T dwarfs obtained 
in various filter systems, due to the strong molecular absorption bands in
their spectra (Stephens \& Leggett 2003). In addition, variability is known 
to occur in T dwarfs (\cf{} Artigau, Nadeau, \& Doyon 2003). 

Thus for present purposes, we adopt the 2MPSC magnitudes on the simple grounds 
that good colour transformations between 2MASS and other photometric systems 
are readily available (Cutri \etal{} 2003) and, in particular, have recently 
been determined for L and T dwarfs explicitly (Stephens \& Leggett 2003). The 
2MPSC magnitudes are given in Table~\ref{tab:finalphot}, along with the 
magnitudes for the individual components, derived using the NACO differential 
measurements from Table~\ref{tab:nacoastphot} and assuming that the NACO and 
2MASS $JHK_s$ colour systems are identical for present purposes. 

\begin{table}
\caption[]{Adopted near-IR magnitudes for the combined \epsindibab{} system 
from the 2MASS Point Source Catalog (Cutri \etal{} 2003) and the derived
individual magnitudes for the two components. The NACO and 2MASS $JHK_s$ 
colour systems are assumed to be identical for this exercise.
}
\label{tab:finalphot}
\begin{center}
\begin{tabular}{lcccc}
\hline
Filter & System    & Combined        & \epsindiba{}    & \epsindibb \\
\hline
$J$    & 2MASS     & \magnit{11}{91} & \magnit{12}{29} & \magnit{13}{23} \\
$H$    & 2MASS     & \magnit{11}{31} & \magnit{11}{51} & \magnit{13}{27} \\
$K_s$  & 2MASS     & \magnit{11}{21} & \magnit{11}{35} & \magnit{13}{53} \\
\hline
\end{tabular}
\end{center}
\end{table}

Finally, a more detailed analysis of the additional optical and IR survey 
data allows us to determine a more refined proper motion for the combined 
\epsindibab{} system: it is now much more consistent with that known for the 
bright primary star, \epsindia{} (Table~\ref{tab:astrom}). The remaining 
difference of $\sim$\,40\,mas/yr is consistent with the expected differential 
motion due to orbital motion of \epsindibab{} around \epsindia: if the 
1459\,AU projected separation corresponds to an orbit lying in the plane 
of the sky, the maximum differential proper motion between \epsindia{}
and \epsindibab{} would be $\sim$\,39\,mas/yr. 

\begin{table}
\caption[]{Proper motions for \epsindibab{} and \epsindia{} in mas/yr.}
\label{tab:astrom}
\begin{center}
\begin{tabular}{@{}lccl@{}}
\hline
Object        & $\mu_\alpha\,\cos\delta$ & $\mu_\delta$       & Source \\
\hline
\epsindibab{} & $+4131\pm 71$            & $-2489\pm 25$      & SMLK03 \\
\epsindibab{} & $+3976\pm 13$            & $-2500\pm 14$      & This paper \\
\epsindia{}   & $+3961.41\pm 0.57$       & $-2538.33\pm 0.40$ & ESA (1997) \\
\hline
\end{tabular}
\end{center}
\end{table}

\section{Spectroscopic observations}
Following the direct imaging, NACO was used in long slit grism mode to obtain
R$\sim$1000 classification spectroscopy in the $H$ band (mode S54\_3\_H,
nominal coverage 1.5--1.85\micron, 6.8\AA/pixel dispersion, S54 camera with
54.3\,mas/pixel). By turning the instrument rotator, both sources were placed
on the slit simultaneously and 12$\times$2 minute exposures were made, 
dithering to different locations along the slit between exposures, for a
total on-source integration time of 24 minutes. 

\begin{figure*}
\centering
\includegraphics[angle=-90,width=18cm]{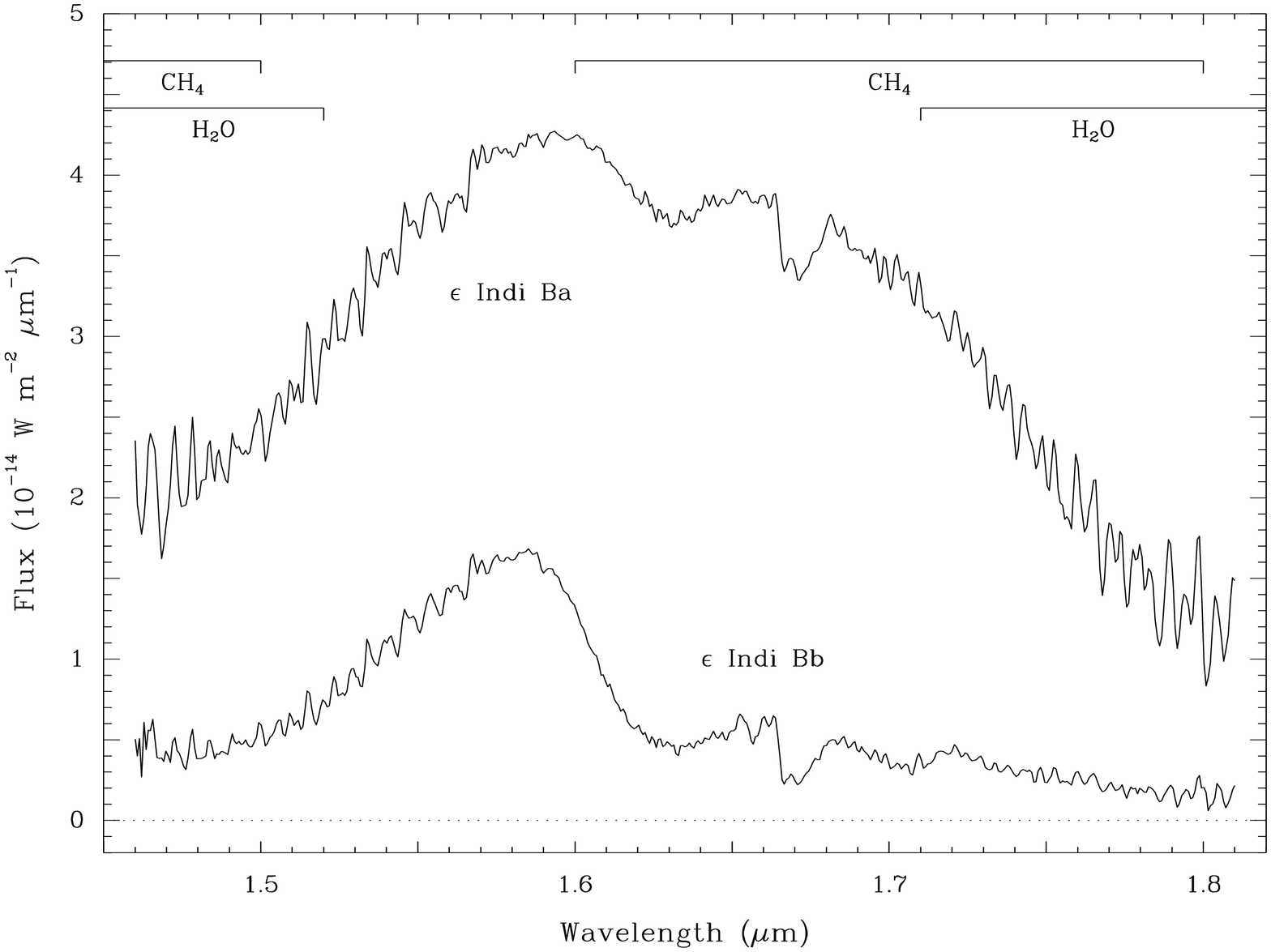}
\caption{$H$ band spectra of \epsindiba{} and \epsindibb. The spectral
resolution is $\sim$17\AA{} FWHM, yielding R$\sim$1000. Flux calibration
was made by convolving the spectrum of \epsindiba{} with the 2MASS $H$ filter 
profile and assuming a 2MASS magnitude of $H$=\magnit{11}{51} as given in 
Table~\ref{tab:finalphot}.
The excellent signal-to-noise of the spectra is seen in the relatively smooth 
1.58--1.62\micron{} range; the `ripples' shortward of 1.56\micron{} and 
longward of 1.72\micron{} are real features, predominantly due to H$_2$O 
and CH$_4$ (\cf{} Geballe \etal{} 2001; Leggett \etal{} 2002; McLean
\etal{} 2003; Cushing, Rayner, \& Vacca, personal communication), but also 
possibly in part due to FeH (\cf{} Cushing \etal{} 2003). The deep double
CH$_4$ absorption trough seen in both sources at 1.67\micron{} is also
seen in the T6 dwarf Gl\,229\,B (Geballe \etal{} 1996), as is the adjacent
absorption feature at 1.658\micron{} seen in \epsindibb.
}
\label{fig:nacospectra}
\end{figure*}

For the spectroscopic observations, the NACO $K$ dichroic was used to send 
full $H$ band flux to the science detector and just the $K$ band flux to the
IR WFS\@. This choice maximised the signal-to-noise in the spectra but
reduced the number of photons available to the IR WFS by roughly 60\%
compared to the imaging. As a result, the adaptive optics correction was
poorer, yielding a spatial resolution of typically 0.3 arcsec FWHM\@.
However, this was nevertheless adequate to ensure well-separated spectra
for the two components of the 0.732 arcsec binary.
 
Observations were also made of
the nearby star HD209552 (G2V) shortly afterwards in order to measure the
telluric absorption. Tungsten-illuminated spectral dome flats were taken in
the same configuration at the end of the night.

Data reduction was standard, employing the IRAF long-slit spectroscopy 
packages. For each source spectral image, several (typically 
three) other images with the sources at different locations were combined to 
make a clean sky image which was subtracted to remove the OH airglow emission. 
The image was then divided by the spectral dome flat. Then returning to the
raw data, the OH lines and the source spectra were traced in order to determine
the geometric transformation which linearised the dispersion, placed the
OH lines horizontally along rows, and the source spectra vertically down
columns. This transformation was applied to all 12 sky-subtracted, flat-fielded
images, which were then aligned and co-added with intensity weighting.

Individual spectra for \epsindiba{} and \epsindibb{} were then optimally 
extracted. By careful measurement of the spatial FWHM along the spectra, it 
was possible to assess the spectral crosstalk as $\sim$2.5\%, \idest{} at
the spatial location of \epsindibb, the flux of \epsindiba{} is reduced to
2.5\% of the flux at its spatial location, and vice versa. At the wavelength
of maximum contrast between the two sources, the brighter source \epsindiba{}
adds roughly 30\% to the flux of the fainter \epsindibb, although more 
typically it is below 10\%. Thus in order to remove most of the crosstalk, an
appropriately scaled version of the \epsindiba{} spectrum was subtracted
from the \epsindibb{} spectrum and vice versa.

Finally, the two source spectra were divided by the spectrum of the atmospheric 
calibrator, similarly reduced and extracted, and then multiplied back by a
template solar spectrum (Maiolino, Rieke, \& Rieke 1996) smoothed to the
resolution of the NACO spectra ($\sim$\,17\AA{} FWHM, R$\sim$1000). Flux 
calibration was achieved using the $H$ band magnitude given for \epsindiba{} 
in Table~\ref{tab:finalphot} and using the 2MASS $H$ filter profile. 

The resulting spectra are shown in Figure~\ref{fig:nacospectra}, with the
major H$_2$O and CH$_4$ absorption bands marked. A more detailed analysis is 
postponed to a future paper, when we should also have higher-resolution spectra 
covering the entire near-IR, but here we simply use the spectra to provide 
spectral classifications using the indices of Burgasser \etal{} (2002) and
Geballe \etal{} (2002). In both systems, the $H$ band contains two indices, 
one measuring the 1.5\micron{} H$_2$O band, the other the 1.6--1.7\micron{} 
CH$_4$ band, and Table~\ref{tab:nacoindices} gives the values and
correspondingly derived spectral types for the two sources. The two Burgasser 
\etal{} (2002) indices and the Geballe \etal{} (2002) CH$_4$ index all give 
relatively consistent spectral types of T1$\pm$0.5 and T6$\pm$0.5 for 
\epsindiba{} and \epsindibb, respectively, while the Geballe \etal{} H$_2$O 
index yields T0 and T4. It is worth noting that this index yields a spectral 
type of T4.5 for Gl\,229\,B, while it is more commonly thought of as $\sim$T6 
based on a broader range of indices (Burgasser \etal{} 2002; Geballe \etal{} 
2002). Thus, for present purposes we assign spectral types of T1 and T6 to 
the two components of the \epsindibab{} system.

\begin{table}
\caption[]{Near-IR $H$ band spectral classification indices for 
\epsindiba{} and \epsindibb{} following the schemes of Geballe \etal{} 
(\cite{geballe02}) and Burgasser \etal{} (\cite{burgasser02}).} 
\label{tab:nacoindices}
\begin{center}
\begin{tabular}{lcccc}
\hline
Index                  & \multicolumn{2}{c}{\epsindiba}
                       & \multicolumn{2}{c}{\epsindibb} \\ 
                       & Value & Spectral Type   & Value & Spectral Type \\ 
\hline
\multicolumn{5}{c}{Geballe \etal{} (\cite{geballe02})} \\ \hline
H$_{2}$O~~1.5\micron{} & 2.01 & T0     & 3.66 & T4 \\
CH$_{4}$~~1.6\micron{} & 1.11 & T1     & 3.27 & T6 \\ \hline
\multicolumn{5}{c}{Burgasser \etal{} (\cite{burgasser02})} \\ \hline
H$_{2}$O\_B            & 0.80 & T1     & 5.29 & T5 \\
CH$_{4}$\_B            & 1.34 & T1     & 6.02 & T6 \\ \hline
\end{tabular}
\end{center}
\end{table}

\section{Revised physical properties}
With the individual magnitudes and spectral types, along with the
accurate distance and relatively well-determined age of the \epsindibab{} 
system, we are now able to estimate their physical parameters, as
summarised in Table~\ref{tab:physical}.

To begin with, we transform the near-IR photometry for the two components
(Table~\ref{tab:finalphot}) from the 2MASS photometric system to the MKO-NIR
system, as the latter is a widely-used standard which is particularly immune
to variations due to the Earth's atmosphere. Also, importantly, Stephens \& 
Leggett (2003) have recently determined a set of transformations 
to the MKO system explicitly for L~and T~dwarfs, which require special 
attention due to their highly-structured atmospheres. The transformations
are parameterised as a function of the spectral type: assuming T1 and T6
for \epsindiba{} and \epsindibb, respectively, as determined from our spectra,
the resulting magnitudes are given in Table~\ref{tab:transformedmags}. Colours 
are also given and are seen to be quite consistent with the corresponding 
spectral types when compared with the compilation of M, L, and T dwarf colours 
in the MKO system plotted in Figure~5 of Leggett \etal{} (2002).

\begin{table}
\caption[]{Transformed near-IR magnitudes and colours for the two components 
of the \epsindibab{} system, using the spectral-type based parameterisation of 
Stephens \& Leggett (2003) to transform between the 2MASS and MKO systems,
and assuming T1 and T6 for \epsindiba{} and \epsindibb, respectively.
}
\label{tab:transformedmags}
\begin{center}
\begin{tabular}{lcrr}
\hline
Filter   & System & 
\multicolumn{1}{c}{\epsindiba{}} & \multicolumn{1}{c}{\epsindibb} \\
\hline
$J$      & MKO    & \magnit{12}{10} &    \magnit{12}{94} \\
$H$      & MKO    & \magnit{11}{56} &    \magnit{13}{31} \\
$K$      & MKO    & \magnit{11}{38} &    \magnit{13}{65} \\
$(J-H)$  & MKO    & \magnit{0}{54}  & $-$\magnit{0}{37} \\
$(H-K)$  & MKO    & \magnit{0}{18}  & $-$\magnit{0}{34} \\
$(J-K)$  & MKO    & \magnit{0}{72}  & $-$\magnit{0}{71} \\
\hline
\end{tabular}
\end{center}
\end{table}

\begin{table*}
\caption[]{Physical parameters for the two components of the \epsindibab{} 
system derived using the COND models of Baraffe \etal{} (2003) covering the 
plausible range of ages (0.8, 1.3, and 2.0\,Gyr) for the system (Lachaume 
\etal{} 1999; SMLK03). See text for a detailed discussion of the derivation 
and the errors in the assumptions, as well as the masses derived in similar
fashion from the Burrows \etal{} (1997) models.
}
\label{tab:physical}
\begin{center}
\begin{tabular}{|ccccc|ccc|ccc|ccc|}
\hline
Source           & M$_K$           & BC$_K$        & M$_{\rm bol}$   &
$\log L/\Lsolar$ & 
\multicolumn{3}{c|}{Mass}          &
\multicolumn{3}{c|}{Radius}        &
\multicolumn{3}{c|}{T$_{\rm eff}$} \\
                 &                 &               &                 & 
                 &
\multicolumn{3}{c|}{\Mjup}         &
\multicolumn{3}{c|}{$R/\Rsolar$}   &
\multicolumn{3}{c|}{K}             \\
                 &                 &               &                 &
                 &
0.8              & 1.3             & 2.0           &
0.8              & 1.3             & 2.0           &
0.8              & 1.3             & 2.0           \\
\hline
\epsindiba{}     & \magnit{13}{58} & \magnit{2}{88}& \magnit{16}{46} & 
$-4.71$          & 
38               & 47              & 57            &
0.096            & 0.091           & 0.086         &
1238             & 1276            & 1312          \\
\epsindibb{}     & \magnit{15}{85} & \magnit{2}{22}& \magnit{18}{07} & 
$-5.35$          & 
22               & 28              & 35            &
0.101            & 0.096           & 0.092         &
835              & 854             & 875           \\
\hline
\end{tabular}
\end{center}
\end{table*}

Next, with the magnitudes in the MKO system, we can use the bolometric 
corrections in that system as determined for late-M, L, and T dwarfs by 
Golimowski \etal{} (2003), which are again parameterised as a function of 
spectral type. This parameterisation yields BC$_K$=\magnit{2}{88} and 
\magnit{2}{22} for the spectral types T1 and T6 of \epsindiba{} and
\epsindibb, respectively. For comparison, the online data of Reid 
({\tt www-int.stsci.edu/$\sim$inr/ldwarf2.html}) give BC$_K$=\magnit{3}{3} 
and \magnit{2}{1} for spectral types T1 and T6, respectively. The differences 
are probably due in part to the differing photometric systems, but also to some
extent to the paucity of T dwarfs with well-determined distances: along with 
the samples of T dwarfs measured in infrared parallax programs (Tinney, 
Burgasser, \& Kirkpatrick 2003), \epsindiba{} and \epsindibb{} will prove 
important additions once their thermal-infrared magnitudes have been 
measured. 

Applying the Golimowski \etal{} (2003) bolometric corrections and the distance
modulus of $-$\magnit{2}{20}, we then derive M$_{\rm bol}$=\magnit{16}{46} and 
\magnit{18}{07} for \epsindiba{} and \epsindibb, respectively. Assuming 
M$_{\rm bol}$=\magnit{4}{69} for the Sun, we then obtain 
$\log L/\Lsolar$=$-4.71$
and $-5.35$ for \epsindiba{} and \epsindibb, respectively. The errors in
this derivation include those in the NACO and 2MASS photometry, the 2MASS--MKO
colour equations, and the distance estimation, but are dominated by the
uncertainty in the bolometric corrections. Following SMLK03, we adopt a
cumulative error of $\pm$20\% in our luminosity determinations.

In SMLK03, we followed the same procedure to this point for \epsindib{} and
then derived T$_{\rm eff}$ by assuming a radius determined from a relationship 
between $R/\Rsolar$ and M$_{\rm bol}$ given by Dahn \etal{} (2002) and modified 
by Reid. That relationship was derived from the evolutionary models of the Lyon 
(Chabrier \etal{} 2000) and Arizona (Burrows \etal{} 1997) groups for L dwarfs 
at $\sim$3\,Gyr and M$_{\rm bol}$=\magap{12}--\magnit{16}{5} and predicts a 
slightly decreasing radius with decreasing luminosity. However, \epsindiba{} 
and \epsindibb{} are younger T dwarfs with M$_{\rm bol}>\magnit{16}{5}$ and,
importantly, lie in a domain where the models predict an {\em increase\/} in 
radius with decreasing luminosity due to electron degeneracy pressure support.  

Therefore here, we use the Baraffe \etal{} (2003) models to extract the radii
for the luminosities and ages appropriate for the \epsindibab{} system, and
thence the effective temperatures. As discussed in SMLK03, Lachaume \etal{} 
(1999) proposed an age of 1.3\,Gyr (with a range of 0.8--2\,Gyr) for 
\epsindia{} based on its rotational properties, and we adopt that age for 
\epsindibab{} here. For the median age of 1.3\,Gyr, the Baraffe \etal{}
(2003) models predict radii of 0.091 and 0.096\Rsolar{} for the luminosities 
of \epsindiba{} and \epsindibb, respectively. Assuming T$_{\rm eff}$=5771\,K 
for the Sun, we then derive T$_{\rm eff}$=1276\,K and 854\,K for \epsindiba{} 
and \epsindibb, respectively. The models predict changes of $\pm$5\% in the 
radii across the 0.8--2\,Gyr age range, yielding corresponding uncertainties 
in the effective temperatures of +30/$-$40\,K for \epsindiba{} and $\pm$20\,K 
for \epsindibb{} (see Table~\ref{tab:physical}).

It should be pointed out that Smith \etal{} (2003) find a significantly higher
effective temperature of $\sim$\,1500\,K for \epsindiba{} based on model
atmosphere fitting of high resolution (R=50,000) near-IR spectra covering
lines of CO, H$_2$O, and CH$_4$. Contamination in their spectra from the 
then-unknown \epsindibb{} appears to be minimal. They note that
spectroscopically-determined effective temperatures are frequently
higher than those calculated using structural models to predict the radius 
as we have done here, although they offer no explanation why this may be the 
case. The degree of the disagreement can be illustrated thus. Smith \etal{} 
(2003) use their effective temperature along with the published luminosity for 
\epsindib{} of SMLK03 to determine its radius: adjusting for the luminosity 
derived here for \epsindiba{} alone, we recalculate that radius as 
0.061\Rsolar, \idest{} considerably smaller than the minimum radius of 
$\sim$0.085\Rsolar{} predicted by structural models for low-mass objects 
at ages 0.8--2\,Gyr. Direct measurements of the radius of \epsindiba{} 
through long-baseline interferometry may help solve this dilemma: at 
3.626\,pc, 0.085\Rsolar{} subtends $\sim$\,0.25 milliarcsec, challenging 
but perhaps not impossible (\cf{} S\'egransan \etal{} 2003).

Finally, we can also use the models to obtain mass estimates for the two
T~dwarfs. For the luminosity of \epsindiba, the Lyon models of Baraffe 
\etal{} (2003) yield masses of $\sim$38--57\Mjup{} for the range 0.8--2\,Gyr, 
with 47\Mjup{} for 1.3\,Gyr. For reference, the Arizona models (Burrows \etal{} 
1997) yield 42--63\Mjup, with 54\Mjup{} for 1.3\,Gyr. For \epsindibb, the 
Lyon models yield 22--35\Mjup, with 28\Mjup{} at 1.3\,Gyr; the Arizona models 
yield 24--38\Mjup, with 30\Mjup{} at 1.3\,Gyr. It is important to note that 
these mass estimates are not significantly affected by the 20\% errors in the
luminosities: the errors are dominated by the age uncertainty. Even then, 
the masses are reasonably well constrained: we adopt the Lyon masses of
47$\pm$10\Mjup{} for \epsindiba{} and 28$\pm$7 for \epsindibb.

\section{Discussion}
It is evident that \epsindiba{} and \epsindibb{} are very special entries
in the growing catalogue of brown dwarfs. They are at a very well-defined 
distance and have a reasonably well-known age, implying that bolometric 
luminosities can be measured accurately and then used to determine masses 
by reference to evolutionary models. The fact that they constitute a binary
system means that the distances and ages are the same for both objects,
making them a powerful differential probe of these models. To date, only
two other T dwarf binaries are known (Burgasser \etal{} 2003) and neither
of these have known ages. 

The proximity of \epsindibab{} to the Sun means that the two components are 
bright and their relatively large angular separation implies that detailed 
physical studies of both sources will be possible, as foreshadowed by the 
results given in the present paper. At the same time, they are also close 
enough to each other physically that their orbits can be measured on a 
reasonable timescale. Assuming masses of 47 and 28\Mjup{} and a semimajor 
axis of 2.65\,AU, a nominal orbital period of 15.5 years is deduced, although 
obviously projection effects, orbital eccentricity, and errors in our mass 
determinations mean that the period could also be either longer or shorter.

The reward for determining the orbit is a model-independent determination of 
the total system mass, in turn placing strong constraints on the evolutionary 
models. Of the other two binary T dwarf systems, one (2MASS 1225$-$2739AB) has 
a possible period of $\sim$\,23 years, while period of the other 
(2MASS 1534$-$2952AB) may be only $\sim$8 years (Burgasser \etal{} 2003): 
however, the latter system is extremely tight with a 0.065 arcsec separation, 
making it very difficult to obtain separate spectra or accurate astrometry. 
Luckily, \epsindibab{} has a a small enough physical separation to help 
determine its orbit fairly quickly and yet it is close enough to the Sun to 
allow both components to be well separated. Indeed, if radial velocity 
variations in the system can also be measured, the individual component 
masses can also be determined. If the orbit were to lie perpendicular to the 
plane of sky, a maximum differential radial velocity $\sim$\,5\kmpers{} would 
be expected. This is readily measurable in principle, although Smith \etal{} 
(2003) have found that \epsindiba{} has a high rotational velocity
($v \sin i$ = 28\kmpers), making it harder in practice. 

In any case, \epsindibab{} will likely prove crucial for an empirical 
determination of the mass-luminosity relation for substellar objects: 
accurate, long-term astrometric and radial velocity monitoring of the pair 
should start as soon as possible.

Finally, it is worth commenting briefly on the announcement of the binary 
nature of \epsindib{} by Volk \etal{} (2003). They suggested that \epsindibb{} 
might either be a brown dwarf companion to \epsindiba{} or, alternatively,
that \epsindibb{} might be a `large planet'. This somewhat provocative latter
hypothesis was, however, unsupported even by their own relatively limited data.
\epsindibb{} is seen to be only $\sim$\magap{1} fainter than \epsindiba{} at 
$\sim$1\micron{} and cursory examination of evolutionary models (\eg{} 
Chabrier \etal{} 2000) reveals that, at $\sim$1\,Gyr, a 10\Mjup{} object would 
be some \magap{4}--\magap{6} fainter than \epsindiba{} at these wavelengths, 
while a 5\Mjup{} object would be $>$\magap{8} fainter. Thus it should have 
been obvious that \epsindibb{} could not be a planet, and indeed, the 
combined imaging and spectroscopy presented in this paper demonstrate clearly 
that \epsindiba{} and \epsindibb{} are `just' brown dwarfs, albeit very 
exciting ones.

\begin{acknowledgements}
We would like to thank Chris Lidman, Markus Kasper, Jason Spyromilio, and 
Roberto Gilmozzi of the VLT on Paranal for their technical and political 
assistance; Isabelle Baraffe for providing versions of the Baraffe \etal{} 
(2003) models for the isochrones appropriate to \epsindibab; Mike Cushing, 
John Rayner, Ian McLean, Sandy Leggett, Denis Stephens, David Golimowski, 
Verne Smith, and Ken Hinkle, for communicating and discussing the results of 
their various papers on T~dwarfs prior to publication. We also thank Isabelle 
Baraffe, Adam Burgasser, Sandy Leggett, Alan MacRobert, Verne Smith, Kevin 
Volk, Ant Whitworth, Hans Zinnecker, and an anonymous referee for comments 
which clarified and sharpened various aspects of the submitted paper. MJM 
thanks the Institute of Astronomy, University of Cambridge, for their valued 
warm hospitality as this paper was started and finished, and Matthew Bate and 
Ian Bonnell for doing what they do best; LMC and BB acknowledge support from 
NASA Origins grant NAGS-12086; and NL thanks the EC Research Training Network 
``The Formation and Evolution of Young Stellar Clusters'' (HPRN-CT-2000-00155) 
for financial support. This research has made use of data products from the 
SuperCOSMOS Sky Surveys at the Wide-Field Astronomy Unit of the Institute 
for Astronomy, University of Edinburgh and from the Two Micron All Sky Survey, 
which is a joint project of the University of Massachusetts and the Infrared 
Processing and Analysis Center/California Institute of Technology, funded by 
the National Aeronautics and Space Administration and the National Science 
Foundation.
\end{acknowledgements}

{}

\end{document}